%Paper: hep-th/9209025
%From: KALYAN%tifrvax.BITNET@cearn.cern.ch
%Date: Tue, 8 Sep 92 22:08 IST +0530

%%%%%%%%%%%%%%%%%%%%%%%%%%%%%%%%%%%%%%%%%%%%%%%%%%%%%%%%%%%%%%%%%%%%%%%%%%
%%%%     PLEASE PROCESS THIS FILE WITH PHYZZX   %%%%%%%%%%%%%%%%%%%%%%%%%%
%%%%%%%%%%%%%%%%%%%%%%%%%%%%%%%%%%%%%%%%%%%%%%%%%%%%%%%%%%%%%%%%%%%%%%%%%%

\def\npb{Nucl. Phys. }

\def\prl{Phys. Rev. Lett. }
\def\prd{Phys. Rev. }
\def\mpla{Mod. Phys. Lett. }

\NPrefs
\def\define#1#2\par{\def#1{\Ref#1{#2}\edef#1{\noexpand\refmark{#1}}}}
\def\con#1#2\noc{\let\?=\Ref\let\<=\refmark\let\Ref=\REFS
         \let\refmark=\undefined#1\let\Ref=\REFSCON#2
         \let\Ref=\?\let\refmark=\<\refsend}

\define\POLB
J. Polchinski, \npb {\bf B331} (1989) 123.

\define\BOK
See e.g. {\it Statistical Mechanics of Membranes and Surfaces}, ed. D. Nelson,
T. Piran and S. Weinberg (World Scientific 1989) and references therein.

\define\P
A. M. Polyakov, \npb {\bf B268} (1986) 406.

\define\PS
J. Polchinski and A. Strominger, \prl {\bf 67} (1991)  1681.

\define\DG
F. David and E. Guitter, \npb {\bf B295} (1988) 332.

\define\PIS
R. D. Pisarski, \prd {\bf D38} (1988) 578.

\define\DNW
S. R. Das, S. Naik and S. R. Wadia, \mpla {\bf A4} (1989) 1033.

\define\DDW
S. R. Das, A. Dhar and S. R. Wadia, \mpla {\bf A5} (1990) 799.

\define\D
F. David, \mpla {\bf A3} (1988) 345.

\define\DK
J. Distler and H. Kawai, \npb {\bf B231} (1989) 509.

\define\PJW
A. Polyakov, Lectures at Jerusalem Winter School, PUPT-1289 (1991)

\def\sg{\sqrt{g}}
\def\rlog{{\rm log}}
\def\sghat{\sqrt{\hat g}}
\def\ghat{{\hat g}}
\def\Rhat{{\hat R}}
\def\ni{\noindent}
\def\intw{\int d^2\xi}
\def\paa{\partial_a}
\def\pab{\partial_b}
\def\pap{\partial_+}

\def\tX{{\tilde X}}
\def\teta{{\tilde \eta}}
\def\half{{1 \over 2}}
\def\gab{g_{ab}}
\def\oalpha{{1 \over \alpha}}
\def\lab{\lambda^{ab}}
\def\tlab{{\tilde{\lambda}^{ab}}}
\def\ghab{{\hat g}_{ab}}

\def\salpha{{\sqrt{\alpha}}}
\def\rz{\rho_0}
\def\meas{[d {\cal M}]}
\def\Liou{S_L (\eta, \ghat)}
\def\piij{\Pi^{i j}}

\def\bz{{\bar z}}
\def\talpha{{\bar{ \alpha}}}
\def\tlambda{{\tilde \lambda}}

\titlepage
\hfill{TIFR-TH-92/48}\break
\title{RENORMALIZATION OF COUPLINGS IN EMBEDDED RANDOM SURFACES}
\author{Sumit R. Das \foot{e-mail: das@tifrvax.bitnet.}
and S. Kalyana Rama\foot
{e-mail: kalyan@tifrvax.bitnet.  \hfill \break
Address after October 1, 1992: School of Mathematics, University of
Dublin, Trinity College, \hfill \break Dublin 2, Ireland.}    }
\address{ Tata Institute of Fundamental Research, Homi Bhabha Road, Bombay
400 005, India}
\abstract{We study the dressing of operators and flows of corresponding
couplings in models of {\it embedded} random surfaces. We show that these
dressings can be obtained by applying the methods of David and Distler and
Kawai. We consider two extreme limits. In the first limit
the string tension is large and the dynamics is
dominated by the Nambu-Goto term. We analyze this theory around a classical
solution in the situation where the length scale of the solution is large
compared to the length scale set by the string tension. Couplings get
dressed by the liouville mode (which is now a composite field) in a
non-trivial fashion. However this does {\it not} imply that the excitations
around a physical ``long string" have a phase space corresponding to an extra
dimension. In the second limit the string tension is small and the
dynamics is governed by the extrinsic curvature term. We
show, perturbatively, that in this
theory the relationship between the induced metric and the worldsheet
metric is ``renormalized", while the extrinsic curvature term receives a
non-trivial dressing as well. This has the consequence that in a generic
situation the dependence of couplings on the physical scale is different
from that predicted by their beta functions.}
\vfill
\endpage

The recent progress in two dimensional gravity and non-critical strings
has considerably enrichened our understanding of a whole class of
field theories on random surfaces. In particular, for random surface
models in which the worldsheet metric fluctuates independently of the
``matter" fields on the surface we have some understanding of how couplings
flow under a change of the physical length scale on the surface.
Consider for example the standard Polyakov string with $d$ matter fields
$X^i$ coupled to the worldsheet metric $\gab$.
Consider a scalar perturbation to the free action $\intw \sg~T(X(\xi))$.
In the conformal gauge $g_{ab} = e^{\phi}\ghat_{ab}$ where $\ghat_{ab}$ is
a fiducial metric, this would become  $\intw \sghat~ e^{\phi}~ T(X(\xi))$.
However, the liouville field $\phi$ has a complicated measure
$[d \phi]$\foot{We denote by $[d \Phi]$ the measure of integration for a
generic field $\Phi$. }
and hence it is convenient to trade $\phi$ for
a scalar field $\eta(\xi)$ having a flat measure \D\DK.
The action for $\eta$ is the standard liouville action
$$ \Liou = { 1 \over { 8 \pi} } \int d^2\xi~\sghat~
(  \ghab \paa \eta \pab \eta - Q \Rhat \eta )  \eqn\liouville  $$
where $Q$ is determined by reparametrization invariance which
translates to Weyl invariance of the full theory of $X^i$ and $\eta$.
For the free theory this
means that the {\it total} central charge is zero, which determines $Q$ to
be ${\sqrt{{25-d \over 3}}}$. In terms of $\eta$ the scalar perturbation
may be written as $\intw \sghat T(\eta(\xi),X(\xi))$ where the
$\eta$-dependence of the {\it dressed} operator $T(\eta(\xi),X(\xi))$ is
determined by reparametrization invariance and given by the equation
\DNW\DDW
$$ [\partial_\eta^2 + Q\partial_\eta + \partial_X^2 + 2] T(\eta,X) +
O(T^2) = 0 \eqn\dress$$
It is crucial to note that the $\eta$-dependence, which controls the flow
of the coupling under changes of the physical length scale, depends on
$T(X)$, whereas the $\phi-$dependence
does not. We shall call such dressings {\it non-trivial}.
A novel feature of the ``renormalization group equations" like \dress\ is
that they
involve higher order derivatives $ \partial_\eta^2 T $
with respect to the scale. This is an effect of quantum
2d-gravity and represents the departure from the semiclassical
expectation that the scale dependence of couplings is given by their beta
functions \foot{For a different approach to higher order RG equations see
\PJW.}. This also
implies that the field $\eta$ has to be interpreted as an extra dimension
in the target space in which the string moves \DNW\DDW\DG\POLB.

For many physical applications, however, it is more appropriate to consider
{\it embedded} random surfaces such as Nambu-Goto strings.
In these models  the worldsheet metric is constrained to be equal to the
induced metric and a good understanding of the flow of couplings is lacking.
At first sight the dressing of operators appears
to be trivial. In fact the action has only
$d$ scalar fields $X^i$ interacting on a flat worldsheet :
there is no quantum gravity to at all! The situation, however,
is not that simple. The Nambu-Goto theories are reparametrization
invariant and $[d X^i]$ is a complicated measure which involves
the induced metric. Indeed it is this feature
that makes the question of dressing non-trivial.

In this letter we address the general question of dressing in two
physically interesting extreme limits. The first limit corresponds to
large string tension where we quantize the theory around a classical
solution whose length scale is large compared with the scale of the string
tension. One such solution is the ``long string" recently discussed in \PS\
and is relevant to string solutions in gauge theories.
We find that the operators in this theory are dressed
non-trivially.
The dependence of the operators on the liouville mode,
which is now a ``bound state" of the $X^i$'s, is labelled by a
continuous parameter.
However because of the non-triviality of the classical solution
the liouville mode does not appear as an extra target space dimension,
unlike the corresponding situation in a Polyakov type string.

In the other limit the microscopic
string tension is very small and the dynamics is
governed by a term proportional to the square of the extrinsic curvature.
This limit is relevant to a description of various kinds of membranes \BOK\
occuring in nature and has been conjectured to be relevant to
QCD strings \P.
The couplings are dressed non-trivially in this case too. More
significantly, the {\it relationship between the induced metric and the
intrinsic metric is also renormalized.}

\ni \underbar{The large string tension limit}

The action of the Nambu-Goto string propagating in a flat Euclidean
d-dimensional space with coordinates $X^i$ is described by the action
$$ S = \mu_0 \intw~\sg \eqn\five$$
where the world sheet metric $ g_{ab} = \paa X^i \pab X^j \delta_{ij} $.
The action \five\ is reparametrization invariant,
and the quantum theory is defined by a path integral with a measure
$[d X^i]$ which respects this invariance. In the conformal
gauge $g_{ab} = e^\phi \delta_{ab}$.
There is, however, a non-trivial Jacobian coming from the change of
variables in the measure. Following the general philosophy of David,
Distler and Kawai \D\DK, it was argued
in \PS\ that this Jacobian results in an additional
liouville contribution
$$ S_L = \beta \intw {\pab (\paa X^i \paa X^i) \pab (\paa X^i \paa X^i)
\over (\paa X^i \paa X^i)^2} \eqn\eight$$
where $\beta $ is a constant.

We quantize the theory around a classical solution $X_0$,
where $ \paa X^i = R e^i_a $ is constant and
$R$ sets the scale. One must have $e^i_a e^i_b = 2 \rho_0 \delta_{ab}$ to
satsify the gauge condition.
Rescaling $X^i \rightarrow {X^i \over R}$ we obtain
$$ S = {1 \over g^2}\intw~[\paa X^i \paa X^i + \beta g^2{\pab
(\paa X^i \paa X^i) \pab (\paa X^i \paa X^i) \over (\paa X^i \paa X^i)^2}]
\eqn\ten$$
where $g = { 1 \over {R \sqrt{\mu_0}} }$ is a dimensionless coupling.
We shall perform a perturbation expansion
in $g$. Expanding around the classical solution
$ X^i = X_0^i + g \tX^i $
the energy momentum tensor becomes \PS\
$$ T_{++} = -{1 \over g} e^i_+ \pap \tX^i - \half \pap \tX^i \pap \tX^i
+ {\beta g} e^i_- \pap^3 \tX^i + O(g^2) . \eqn\twelve$$
$T_{++}$ satisfies virasoro algebra with central charge
$ c = d + 12 \beta + O(g) $.

To see the equivalence of the above quantisation to that of two
dimensional quantum gravity a la \D\DK\ we note that the action \eight\ is
the liouville action with the liouville field $\phi$ constrained to satisfy
$e^\phi = \paa X^i \paa X^i$. Following standard procedure we trade $\phi$
with a scalar field $\eta$ which has a flat measure.
We introduce a lagrange multiplier
$\lambda(\xi)$ to impose the constraint. In usual conformal field theory
strict conformal invariance is equivalent to the requirement of Weyl
invariance of the theory coupled to some fiducial two dimensional metric.
We thus introduce a fiducial metric $\ghat_{ab}$ and consider the action
$$ S = { 1 \over g^2} \intw~\sghat[(1 + \lambda(\xi))\ghat^{ab} \paa X^i
\paa X^i - \lambda~e^{\eta} +g^2 \beta(\ghat^{ab}\paa \eta \pab\eta -
{\hat R} \eta )] \eqn\fourteen$$
where we have already scaled various fields to bring the coupling out of
the entire action. The curvature term is the standard coupling of the
liouville field and gives the correct energy momentum tensor \foot{One can
also view the action \fourteen\ as that arising from the original
Nambu-Goto theory where the constraint which equates the worldsheet metric
to the induced metric is imposed by a largrange multiplier $\lab$ and a
conformal gauge $g_{ab} = e^\phi~\ghat_{ab}$ is picked.
The change of measure from $\meas_g$
defined with respect to $g_{ab}$ to $\meas_{\ghat}$ would give rise to the
standard Liouville action if the original theory was exactly conformally
invariant. In our case the theory is interacting, but we shall see that to
lowest order the liouville action is sufficient. The action \fourteen\
arises if we integrate out $\lambda_{\pm \pm}$ and decide to impose the
resulting constraints on the space of states. To lowest order in $g$ these
constraints are identical to $T_{\pm \pm}$ (eqn. \twelve)}.
We will always work on a worldsheet of spherical topology, so that we
can take $\ghat_{ab} = e^{\sigma} \delta_{ab}$.
The classical solution is now
$$ \paa X_0^i = e^i_a = {\rm constant}~~~~e^i_a e^i_b = 2 \rho_0
\delta_{ab} ~~~~\lambda_0 = 0~~~~~~~\eta (\xi) = \eta_0 - \sigma (\xi)
\eqn\fifteen $$
where $\eta_0 = \rlog \rho_0$ is constant. Expanding around \fifteen\
$$ X^i = X^i_0 + g \tX^i~~~~~\lambda = g \tlambda~~~~~~
\eta = \eta_0 - \sigma + g \teta \eqn\fifteena$$
and changing to the new variables
$Y^i (p) = \tX^i(p) - {i (e^i \cdot p) \over p^2} \tlambda(p) $ and
$Z(p) = \tlambda (p) + \teta (p) $
in momentum space, the action in \fourteen\ becomes
$$ S^{(2)} = \int d^2p~\half[ p^2 Y^i(p)  Y^i(-p) - \rho_0 Z(p)
Z(-p) + (g^2\beta~p^2 + \rho_0) \teta(p) \teta (-p)] . \eqn\seventeen$$
In the following we will need the $X^iX^j$ propagator to $O(g^2)$.
Remarkably, upto this order the radiative corrections of $O(g^2)$ coming from
the interaction terms cancel and the propagator is given entirely by the
quadratic action. We list below some
propagators:
$$ \eqalign{ & <\tX^i (p) \tX^j(-p)> = {1 \over p^2}[\delta^{ij} -
{\beta g^2 (e^i_a p_a)(e^j_b p_b) \over \rho^2}] + O(g^4) \cr &
<\tX^i (p) \teta (-p)> = -{ie^i_a p_a \over \rho_0 p^2} +
O(g^2) . }    \eqn\eighteen $$

The $O(1)$ contribution to the central charge may be calculated
by integrating out the fields using the action \seventeen\ upto $O(g)$
with a regulator which respects reparamaetrization invariance
of the fiducial metric theory and calculating the coefficient of the term
$\intw (\partial \sigma)^2$.
It is then clear from \seventeen\ that the
fields $Y^i$ contribute $d$, the fields
$Z$ and $\teta$ contribute zero and the ghost fields
contribute $- 26$ to the central charge.
\foot{The change of variables from $(\tX,\tlambda,\teta) \rightarrow
(Y,Z,\eta)$ does not produce a nontrivial Jacobian}
However, substituting the
expansions \fifteena\ in \fourteen\ also leads to
a {\it classical} contribution of $ 12 \beta$ to the central charge.
The reparametrization invariance of the original
theory, or equivalently the Weyl invariance of
the theory of $\tX^i,\teta,\tlambda$
on a fixed background metric $\ghat_{ab}$, implies that
$ 12\beta = 26 -d $.
Unlike the Polyakov string the $26$ does not become $25$ since
the field $\teta$ is not a propagating field to $O(1)$.
In higher orders of $g$ the ``liouville action" is more
complicated but presumably determined by the same principle.

How are the operators dressed ?
Consider adding to the action a term which, in the conformal gauge, is
$ S_I = \gamma \intw \sghat e^\phi T(X^i)$.
The question of dressing now
is : what is $S_I$ in terms of the field
$\eta$ ? To lowest order in the coupling $\gamma$ this may be computed by
demanding reparametrization invariance which requires
the expectation value $<S_I>$ to be independent of $\ghab$.
Let us begin with an ansatz for the form of $S_I$, viz. $S_I \sim \gamma
\intw\sghat ~e^{\alpha \eta}~T(X)$.

Consider first $T(X) = e^{ik_iX^i}$ where $k$ is $O(1)$ implying that
the momentum excitations are
small and the {\it backgrounds
vary slowly in space}.
We expand the fields as in \fifteena\ and use the standard
propagator in curved space ( $ \ghab = e^{\sigma} \delta_{ab} $ )
$$ \int d^p{e^{ip \cdot (z_1 -z_2)} \over p^2} \rightarrow -{\half} \rlog
[ m_0^2 \vert z_1 - z_2 \vert^2
+ a^2 e^{- \sigma({z_1 + z_2 \over 2})}]
\eqn\nineteen$$
where $m_0$ is an infrared cutoff and $a$ is the lattice spacing
and obtain the following $\sigma$-dependences
$$ <\tX^i (z) \tX^j (z)> = \half \sigma(z) + O(g^2) ~~~~~~<\tX^i (z) \teta (z)>
= {1 \over 4} e^i_a \paa \sigma(z) .  \eqn\twenty$$
These expressions will suffice to calculate $<S_I>$
to $O(g^2)$. The $\sigma$-dependence is
$$ <S_I> \sim \intw e^{\alpha \eta_0 + i k \cdot X_0}~ e^{(1 -
\alpha)\sigma} ~[ 1 - {1 \over 4} g^2 k^2_{tr}~\sigma (\xi) + O(g^4)]
\eqn\twentyone$$
where $k^i_{tr} = \piij k^j $
and $\piij = \delta^{i j} - { 1 \over \rz } e^i_a e^j_a $  is the
projection operator. Thus for $\sigma$-independence we must have
$$ \alpha = 1 ~~~~~~~~k^2_{tr} = 0   \eqn\twentytwo$$
which is expected classically. Hence
we conclude that for the {\it slowly varying perturbations the dressing
is trivial.} Consequently no extra degree of freedom appears.

For $k \sim O({1 \over g})$, the momenta of excitations are
at the scale of the string tension.
It is now more convenient to rescale $ k \rightarrow {k \over g} $ and
the new $k$ is of order unity. The result for $<S_I>$ now reads
$$ <S_I> = \intw~e^{\alpha \teta + i {k \over g} \cdot X}~ e^{[1-(\alpha +
\half k^2)]\sigma}~[1 + \half {i \alpha g \over \rho_0} (e^i_a k^i) \paa \sigma
(\xi) + O(g^2)] \eqn\twentythree$$
This shows that $S_I$ as it stands cannot be dressed consistently. We find
that the following modified $S_I$
$$ \intw \sghat[ e^{\alpha \eta + ik \cdot X}
+ \half \alpha g^2 e^{(\alpha-1) \eta
+ i {k \over g} \cdot X}~{\hat \nabla}^2 \eta] \eqn\twothreea$$
leads to $\sigma$-independence upto $O(g)$ provided
$ \alpha + \half k^2 = 1 $
\foot{Note that though the two terms in \twothreea\ have
different powers of $e^\eta$
they correspond to the same scaling behavior. This is
because the original action is invariant under the simultaneous
transformations
$ X \rightarrow \kappa X$, $ \eta \rightarrow \eta + \rlog \kappa^2$ and
$g \rightarrow \kappa g $.
The extra power  of $g^2$ in the second term of \twothreea\
compensates for the different power of $e^\eta$.}.
The dressing is now {\it nontrivial} and depends on $k$.
The equation which determines the dressing differs from that in a
Polyakov theory by the absence of
$\alpha^2$ terms. This happens because the liouville mode $\eta$,  which
is now actually a composite of the $X^i$'s, is not a massless field.
Consequently the flow equations for the coupling $\gamma$ are first order.

It is instructive to verify this in a
different approach. Integrating out the lagrange multiplier, of course,
leads to the theory of $X^i$'s given by \ten.
In this theory, now formulated in flat space, a perturbation to the action
must correspond to a $(1,1)$ vertex operator. We shall verify this to
order $g$ using the energy momentum tensor given by \twelve.
To do this we have to construct the vertex operator to $O(g^2)$.
This means that we have to replace $\eta$ in
$e^{\alpha \eta + i{k \over g} \cdot X}$ by the solution of
$ e^{\eta} = \rho_0~e^{g\teta} = \paa (X^i_0 + g\tX^i)\paa (X^i_0 +
g\tX^i)$ to the relevant power of $g$.
This yields
$$ V(w) = \rho_0^\alpha~e^{i{k \over g} \cdot X_0}~e^{ik \cdot \tX}~
[1 + { {\alpha g} \over \rho_0} (e^i_a \paa \tX^i)
+ { {\alpha g^2} \over {2 \rho_0} } v_1 + O(g^3)] \eqn\twentyfive$$
where $ v_1 = \paa X^i \paa X^i + {\alpha - 1 \over \rho_0}
(e^i_a \paa X^i)^2 $.
The operator product expansion now reads
$$T_{++}(z) V(w) \sim
{ig\alpha k^ie^i_{-} V(w) \over 2 \rho_0 (z-w)^3}
+ {(\half k^2 + \alpha) V(w) \over (z-w)^2}
+ {\partial V(w) \over (z-w)} + O(g^2) \eqn\twentyseven$$
The cubic pole exactly corresponds to
the term involving $\partial \sigma$ in the curved space calculation. This
is quite common in conformal field theory.
It is easily seen that adding another term
$ v_2 = {i \alpha g^2 \over \rho_0^2} \rho_0^\alpha e^{i { k \over g} \cdot
X_0}~ (e^i_-k^i)(e^j_-\pap^2 \tX^j) $
to the vertex operator $V(w)$ in \twentyfive\
cancels the cubic pole in \twentyseven\
and the new $V(w)$ is a $(1,1)$ primary provided
$\alpha + \half k^2 = 1$. $v_2$ is precisely the second term in
\twothreea.

%The main lesson of this exercise is that while for perturbations with
%small wave number the liouville mode does not play any role (i.e. the
%dressings are trivial), the same is not true for high wave-vector modes.
%In some sense the theory perceives an extra degree of freedom at high
%energies.

What does this mean for the spectrum of excitations around the long string
considered in \PS\ ? It must be emphasized that while we derived
the above dressings by considering perturbations around a specific
classical solution, the result is independent of the particular solution
we have expanded around. However, the spectrum of excitations around a
given classical solution clearly depends on the properties of the
classical solution in question. Consider for example a string solution
given by $X^i_S(z,\bz) = e^i_+ z + e^i_- \bz$ where $(z,\bz)$ are
the complex coordinates on the {\it sphere}. Expand around this solution
$ \partial_z X^i(z,\bz) = \partial_z X^i_S (z,\bz) + g \sum_{n} \alpha^i
(n) / z^{n+1} $.
The ground state $|0>_S$ is annihilated by $\alpha^i (n)$ for $n > 0$.
The operator products derived above then imply that the state
$|\alpha,k> = V(0)|0>_S$, with $V(z)$ discussed above, satisfies the
physical state conditions $L_0 |\alpha,k> = {\bar L}_0 |\alpha,k> =1; L_n
|\alpha,k> = 0 (n >0)$ provided $\alpha + \half k^2 = 1$. The excitations
seem to live in one extra dimension, the extra momentum being identified
with $\alpha$.

The main point, however, is that {\it this classical solution does not
describe a physical string}. A physical string should have a {\it real}
$X^i(\sigma, \tau)$ where $\sigma, \tau$ are the real coordinates on the
cylinder. It is easy to check that when the solution $X^i_S$ above is
conformally transformed to the cylinder and the euclidean time is
continued back to real $\tau$, the configuration obtained is no longer
real \foot{In fact the reality condition translates to $(X^i(z,\bz))^* =
X^i(\bz^{-1}, z^{-1})$ in terms of the complex coordinates on the sphere.}.

A physical long string corresponds to a solution $X^i_L(\sigma,\tau) =
e^i_+ (\sigma + \tau) + e^i_- (\sigma - \tau)$. The conformally
transformed and the suitably analytically continued solution corresponds
to the following configuration on the sphere : $X^i_L(z,\bz) = e^i_+
\rlog z + e^i_- \rlog \bz$ which obviously satisfies the reality
condition. The long string vaccuum $|0>_L$ is annihilated by
$\talpha (n)$ for $n > 0$, where
$ \partial_z X^i(z,\bz) = \partial_z X^i_L (z,\bz) + g \sum_{n}\talpha^i
(n)/ z^{n+1} $. It may be now easily checked that a state $V(0)
|0>_L$ is physical only if
$$ \alpha = 0~~~~~(e^i_+ - e^i_-)k^i = 0~~~~~{1 \over g} e^i_+ k^i +
\half k^2 = 1 \eqn\twothreea$$
Since $\alpha$ is zero {\it independently} of the value of the momentum,
there is no extra dimension.

We could have gone through the same exercise for a standard non-critical
Polyakov string. In that case we would have obtained
$(e^i_+ - e^i_-)k^i = 0$ and ${1 \over g} e^i_+ k^i + \half k^2
+ { 2 \over Q^2} \alpha (\alpha + Q^2) = 1 $.
(Note that the definition of $\eta$ we are using differs from the standard
definition by a factor of $-{2 \over Q}$.)
In this case the liouville acts as an extra dimension as usual. The reason
why it does not for the Nambu-Goto string is that the field $\eta$ is not
independent of the field $X^i$. Thus the allowed values of the
$\eta$-momentum $\alpha$ is sensitive to the particular classical solution.

\ni \underbar{The rigid String}

In the limit of vanishingly small surface tension, the dynamics is
controlled by an extrinsic curvature term. For simplicity we shall work by
setting the string tension to be exactly zero. The results can be,
however, generalized to non-zero string tension. The action is
$$ S = {1 \over {2 \alpha} } \intw [ \sg (\nabla^2X^i)^2
+ \lab (\paa X^i \pab X^i - \gab)] \eqn\eone$$
where $\lab$ is the lagrange multiplier which imposes the constraint that
the induced metric is equal to the worldsheet metric.

This model has been studied over the past few years in some detail
\P\BOK\DG\PIS. The
most important result is the fact that the coupling $\alpha$ is
asymptotically free. As in other aymptotically free models, this implies
that either (i) $\alpha$ hits a non-trivial fixed point leading to a
nontrivial continuum limit in which large curvature fluctuations are
supressed, or (ii) the coupling keeps growing in the infrared. In case (ii)
one expects that $\lab$ acquires
a non zero expectation value with small fluctuations so that the
constraint is effectively absent while in the deep infrared the coupling
$\alpha$ grows large rendering the first term irrelevant. The model then
effectively reduces to a Polyakov string with $d$ scalar fields
interacting with two dimensional gravity \P. At present it is not known which
of the above scenarios is correct. Large-$d$ expansion vindicates (ii)
while there is numerical evidence for a
second order phase transition that may be consistent with (i).

None of the existing studies of this model has addressed the {\it
renormaliztion} of the coupling. In fact it is natural to expect
that the renormaliztion
of couplings in this theory will be non-trivial. After all if the scenario
(ii) is correct, the theory becomes the Polyakov string in
the infrared limit which has a $(d+1)$
dimensional target space. In fact, as we shall see below, the non-trivial
renormaliztion of the couplings determines the scaling behavior of the
model.

As before we work in
the conformal gauge $\gab = e^\phi \ghab$ and trade the liouville mode
$\phi$ for the standard scalar field $\eta$. We start with the following
ansatz for the dressed action
$$ S = {1 \over {2 \alpha} }
\intw [ \sghat G(\eta) e^{-\eta}({\hat \nabla}^2 X^i)^2
+ \lab ( \paa X^i \pab X^i - H(\eta) e^\eta \ghab ) ] +
\Liou \eqn\etwo$$
where $\Liou$ is given to the lowest order in $\alpha$ by \liouville\ and
$G(\eta)$ and $H(\eta)$ are arbitrary functions. The function $H$ in
\etwo\ modifies the relation between the world sheet and the induced
metrics to $ \paa X^i \pab X^i = H(\eta) e^\eta \ghab $.
We will determine below
the coefficient of $S_L$ and the functions $G$ and $H$ to the lowest
order in $\alpha$. In higher orders the action will get modified.

In the following it suffices to take
$\ghab = e^\sigma \delta_{ab}$. We expand around
the classical solution
$ X^i = X_0^i + \salpha \tX^i$,
$\eta = \eta_0 - \sigma + \salpha \teta$ and $\lab = \salpha \tlab $
where $\paa X^i_0 = e^i_a $  is constant and
$e^{\eta_0} = \rho_0 H(\eta_0) $. We further
define $\zeta = \tlambda^{aa} $. Expanding the function $G(\eta)$ and
ignoring $O(\sigma^2)$ terms
$$ G(\eta) = G^{(0)}(\eta_0) + \salpha G^{(1)}(\eta_0) \teta + {\alpha
\over 2} G^{(2)}(\eta_0) \teta^2 + \cdots \eqn\efour$$
where
$ G^{(n)}(\eta) \equiv (\partial_\eta^n - \sigma \partial_\eta^{n+1})G
(\eta) $, and similarly for $H(\eta)$,
the quadratic part of the action becomes
$$ S^{(2)} = {1 \over 2}\intw~
[ G^{(0)} {1 \over \rho_0} (\partial^2 \tX^i)^2
+ 2 \tlab e^i_a \pab \tX^i - \rho_0 ( H^{(0)} + H^{(1)} ) \zeta \teta ]
\eqn\esix$$
while the cubic and quartic parts are
$$\eqalign{S^{(3)} = {\salpha \over 2} \intw~[&-{1 \over \rho_0}
(G^{(0)}-G^{(1)}) \teta (\partial^2 \tX^i)^2
+ \tlab \paa \tX^i \pab \tX^i
\cr &
- \half ( H^{(0)} + 2 H^{(1)} + H^{(2)} )
\rho_0 \zeta \teta^2 ]} \eqn\eseven$$
$$\eqalign{S^{(4)} = {\alpha \over 2}\intw[{1 \over 2
\rho_0}(G^{(0}- & 2G^{(1)}+G^{(2)}) \teta^2 (\partial^2 \tX)^2 \cr &
-{1 \over 6}\rho_0(H^{(0} + 3 H^{(1)} + 3 H^{(2)} + H^{(3)})\zeta \teta^3]}$$
Clearly $\partial_\eta G$ and $\partial_\eta H$ are at least of
$O(\alpha)$. Hence we can
calculate the propagators to $O(\alpha)$ by neglecting $H^{(1)}$ and
setting $G^{(0)} = G(\eta_0)$  and $H^{(0)} = H(\eta_0)$ in
$S^{(2)}$. A similar comment applies to $S^{(3)}$ and $S^{(4)}$.
{}From now on the functions $G$, $H$ and their derivatives  are assumed
to be evaluated at $\eta_0$ and we will not write their arguments
explicitly.  Defining the fields
$f^a$ and $Y^i$ by
$i({\hat \nabla})^2 f^a = {\hat \nabla}_b\tlab$ and
$Y^i(p) = \tX^i(p) - {i \rho_0 \over {G p^2}  }e^i_a f^a(p)$
in momentum space, the action $S^{(2)}$ becomes
$$S^{(2)} = \half \int d^2p [{G \over \rho_0}
p^4 Y^i(p) Y^i(-p) + { {\rho_0^2 H} \over G} f^a(p)f_a(-p)
- \rho_0 H \zeta (p) \teta (-p) ] \eqn\eten$$

To compute the central charge, $c$, we note that the Jacobian for the change
of variables $(\tX^i,\tlab,\teta) \rightarrow (Y^i, f^a,\zeta,\teta)$ is
${\rm det}~(\partial^2)$ which contributes $- 2$ to $c$.
{}From \eten\ it follows that the only contribution to $c$ is from the
field $Y^i$ and is equal to $2 d$
because the kinetic term has $p^4$ instead of $p^2$.
As in the long string example, there is a
classical contribution of  $12 \beta$ from the liouville part of the
action and thus the total central charge is
$(2 d - 2 + 12 \beta)$ \foot{In the large string tension limit,
the relevant change of variables are $i({\hat \nabla})^2 f^a
= {\hat \nabla}_b\tlab$ and
$Y^i(p) = \tX^i(p) - i \rho_0 e^i_a f^a(p)$. Then the action has a $p^2 Y^i
Y^i$ and a $p^2 f^a f^a$. Thus the central charge conributions are $d$
from the $Y^i$ fields, 2 from the $f^a$ and $-2$ from the jacobian,
yielding a total central charge $d + 12\beta$. This is exactly what we
obtained in the previous section.}.
This may be also verified by computing the
energy momentum tensor of the theory and directly evaluating
$<T_{++}(z)T_{++}(w)>$.
Thus for $\sigma$ independence of the action to $O(1)$
we must have $c = 26$ or equivalently $\beta = {14 - d \over 6}$.
As expected $\beta \sim O(1)$. Consequently we can ignore $\Liou$
in performing one loop calculations.

The non zero propagators for the various fields
$$\eqalign{&
<\tX^i(p) \tX^j(-p)> = {\rz \over {G p^4} } \piij  ~~~~~~~~
<\tX^i(p)f^a(-p)> = {i e^i_a \over {\rz H p^2} } \cr  &
<f^a(p) f^b(-p)> = { {G \delta^{ab}}  \over {\rz^2 H} }~~~~~~~~~~~~~
<\zeta(p) \teta (-p)> = {2 \over {\rz^2 H} }   } \eqn\eeleven$$
where $\piij$ is given after equation \twentyone.

It is important to note that to the lowest order
$<\teta (p) \teta (-p)> = 0$, $\teta$ being the liouville field. This
will have important implications for the dressing. The physical reason for
this lack of correlation is that
the world sheet metric is constrained to be
proportional to the induced metric.

The $\sigma-$dependence of the action arises from
(i) the functions $G^{(0)}$ and $H^{(0)}$
and (ii) the one loop diagrams. The latter only involves the
{\it divergent} parts of the one loop diagrams since to this order
the quadratic and higher terms in $\sigma$ do not
contribute. Using standard
dimensional regualrization and replacing $\intw$ by
$\intw~e^{ - \epsilon \sigma}$, we find the relevant counterterms to be
$$ S_{\rm ct} =
{ \alpha \over {2 \pi \epsilon} } \intw~e^{- \epsilon \sigma}
[-( {1 \over {G H} } )  { G \over \rz } (\partial^2 X^i)^2
+ ({d-2 \over { 2 G H } }) \rz H \zeta \teta] \eqn\etwelve$$
The divergent pieces are absorbed by renormalizing
the coupling $\alpha$ to $\alpha_R$ in the standard way,
for example, as in \P\BOK, leading to a beta function $\beta(\alpha)
\equiv a{\partial \alpha \over \partial a} = {d \alpha^2 \over 4 \pi}$.
The total one loop effective action to $O(\alpha)$ becomes
$$ S_{eff} = {1 \over {2 \alpha_R} }
\intw~[ Z_{xx} {G \over \rz} (\partial^2 X^i)^2
+ 2Z_{fx} \lab e^i_a \pab X^i - Z_{\zeta \eta} \rz H \zeta \eta]
\eqn\ethirteen $$
where
$Z_{xx} = 1 - ( {G' \over G } + {\alpha \over { 4 \pi G H} } ) \sigma $,
$Z_{fx} = 1 $ and
$ Z_{\zeta \eta} = 1 + ( { {H' + H''} \over H}
+ {\alpha (d-2) \over {8 \pi G H} } ) \sigma $. The terms involving
derivatives of the functions $G$ and $H$ come from the $O(\alpha)$ terms
in $S^{(2)}$ (equation \esix) while the others come from the
counterterms.
The $\sigma$ independence of the effective action above is obtained by
demanding
$$G'(\eta) + {\alpha \over { 4 \pi H} } =
H' + H'' + \alpha { {d - 2} \over {8 \pi G} } = 0  . \eqn\efourteen$$
Note that
\efourteen\ implies the equation
${\cal G}' + \alpha { d \over {8 \pi} } = 0  $ for the product
${\cal G} = G (H + H')$.

The second equation in \efourteen\ renormalizes the relation between the
world sheet and the induced metrics. To obtain the theory entirely in
terms of the fields $X^i$ one has to substitute $e^\eta = H^{-1}~\paa X^i
\paa X^i$.
In other words the renormalized scale factor of the
surface is given by a field $\psi (\xi)$ defined by $e^\psi =
e^\eta~H(\eta)$.
A constant shift of $\psi(\xi)$ then corresponds to a
scale transformation $X \rightarrow \kappa X$. The extrinsic curvature
term is then $\oalpha \intw (G H) e^{-\psi} ({\hat \nabla}^2~X)^2$.
The physical coupling is thus $\beta(\psi) \equiv {\alpha \over
G(\psi)H(\psi)}$. The equation \efourteen\ then implies that to the lowest
order the equation determining the scale dependence of $\beta(\psi)$ is
given by \foot{In the lowest order calculation we are doing, one can
consistently replace $\partial / \partial \eta$ by $\partial / \partial
\psi$. }
$$ {\partial \beta \over  \partial \psi} = {d \over 8 \pi} \beta^2 +
{H'' \over H}\beta \eqn\efifteen$$
The first term on the right hand side is the beta function for the
coupling. The presence of the second term shows that the dependence on the
physical scale is not entirely given by the beta function. We expect this
to be a generic feature of models of embedded random surfaces
\foot{In our particular case of
interest there are two solutions of the equations \efourteen\ one of which
indeed has $H''=0$, while the other has $H''\sim \alpha {\rm exp}~(-\psi)$.
Unlike
the standard examples of $c < 1$ matter coupled to gravity there is no
correspondence principle which eliminates one of the solutions and at this
point it is unclear which solution has to be picked out.}.

Finally we make some remarks about the behavior of the couplings beyond
perturbation theory. Much of the above considerations
depend on the fact that the lagrange multiplier field $\lab$
has a zero expectation value. Non-perturbatively
one may expect $\lab$ to
acquire a non-zero expectation value : one immediate consequence of this
is that the two point function of the ``liouville" field $\eta(\xi)$ is
non-zero. This happens, for example in the large-$d$ limit \DG\PIS.
At $d=\infty$ the dressing of the extrinsic curvature term
is similar to our one loop
perturbative result. However, the non-trivial dressing will modify the
${1 \over d}$ corrections. Preliminary investigations show that the
non-trivial dressing does not change the overall qualitative picture of \DG.
In particular, when
{\it renormalized} string tension is tuned to zero, the fluctuation of the
metric around the large-$d$ saddle point, $\teta(\xi)$, behaves like a
massless field for length scales much
larger than the length scale of the dynamically generated $<\lab>$.
Furthermore, the fluctuations of $\lab$ around its saddle point value are
suppressed, so that effectively {\it the liouville mode becomes an
independent field}.
Owing to asymptotic freedom of $\alpha$ the $X$- propagator also becomes
${ 1 \over p^2}$. The
dressing of operators then assumes a form similar to that in the Polyakov
string and the liouville mode results in an extra physical dimension. We
shall discuss this case in detail in a future communication.

\centerline{\underbar{Acknowledgements.}}

We would like to thank G. Mandal, N. Seiberg, A. Sen, A.
Sengupta, S. Shenker, A. Strominger, S. Wadia and especially
A. Dhar for discussions and P. Nelson for a correspondence.
S.R.D. would like to thank the International Center for Theoretical Physics,
Trieste for hospitality during the early stages of the work.
\refout

\end